\documentclass[aps,twocolumn,groupedaddress]{revtex4}
\usepackage{amsmath}
\usepackage{amssymb}
\usepackage{epsfig}
\usepackage{tabularx}
\usepackage{rotating}
\usepackage{color}
\usepackage{pslatex}

\def\cmm1{cm$^{-1}$}
\def\kjmol{kJ$\cdot$mol$^{-1}$}
\def\etal{{\it et al.}}
\def\fmthree{Fm$\bar{3}$}
\def\poneht{P1$_{HT}$}
\def\pbone{P$\bar{1}$}
\def\pbonelt{P$\bar{1}_{LT}$}
\def\pboneht{P$\bar{1}_{HT}$}
\def\pbthree{P$\bar{3}_{HT}$}
\def\ponep{P1$^\prime_{HT}$}
\def\csixty{C$_{60}$}

\def\csixtx{C$_{60}$~$\cdot$~2~CHX$_{3}$}
\def\csixcl{C$_{60}$~$\cdot$~2~CHCl$_{3}$}
\def\csixbr{C$_{60}$~$\cdot$~2~CHBr$_{3}$}
\def\clform{CHCl$_{3}$}
\def\brform{CHBr$_{3}$}

\def\dE{{$\Delta E$}}

\def\Efermi{{$E_F$}}

\def\xray{X--ray}
\def\bda{\boldsymbol a}
\def\bdb{\boldsymbol b}
\def\bdc{\boldsymbol c}
\def\bdk{\boldsymbol k}
\def\bdr{\boldsymbol r}
\def\bdG{\boldsymbol G}
\def\bdR{\boldsymbol R}

\newcommand{\E}[1]{\times 10^{#1}}

\newcommand{\tbn}[1]{$^{\bf #1)}$}

\newcommand{\dist}[1]{$R(${#1}$)$}

\newcolumntype{R}{>{\raggedleft\arraybackslash}X}
\begin{document}

%


\title{
Crystal Structures and Electronic Properties of
Haloform-Intercalated \csixty
}

\author{
Ren\'e Windiks, Andreas Bill and Bernard Delley
}
\affiliation{
Paul Scherrer Institut, Condensed Matter Theory Group,
CH--5232 Villigen PSI, Switzerland
}
\email{rene.windiks@psi.ch}
\author{
V.~Z.~Kresin
}
\affiliation{
University of California, Lawrence Berkeley Laboratory,
Berkeley CA 94720, USA
}

\date{\today}

%
\begin{abstract}

Using density functional methods we calculated structural and electronic
properties of bulk chloroform and bromoform intercalated \csixty,
\csixtx\ (X=Cl,Br).
Both compounds are narrow band insulator materials with a gap 
between valence and conduction bands larger than 1~eV.
The calculated widths of the valence and conduction bands are $0.4$--$0.6$~eV 
and $0.3$--$0.4$~eV, respectively.
The orbitals of the haloform molecules overlap with the $\pi$ orbitals of the 
fullerene molecules and the $p$-type orbitals
of halogen atoms significantly contribute to the valence and conduction
bands of \csixtx.
Charging with electrons and holes turns the systems to metals.
Contrary to expectation, 10 to 20~\% of the charge is on the haloform 
molecules and is thus not completely localized on the fullerene molecules.
Calculations on different crystal structures of \csixcl\ and \csixbr\
revealed that the density of states at the Fermi energy are
sensitive to the orientation of the haloform and \csixty\ molecules.
At a charging of three holes, which corresponds to the superconducting phase of 
pure \csixty\ and \csixtx, the calculated density of states (DOS) at the Fermi 
energy increases in the sequence DOS(\csixty) $<$ DOS(\csixcl) $<$
DOS(\csixbr).

\end{abstract}
%


\maketitle

%
\section{Introduction}
\label{sec:intro}

This paper is concerned with structural and electronic properties of solid
\csixty\ intercalated with haloform molecules, CHX$_3$ (X=Cl,Br).
These compounds have been studied recently by \xray\ diffraction
\cite{jansenwaidmann14-1995,collinsfoulkesbondklinowski1-1999,
dinnebiergunnarssonbrummkochhuqstephensjansen-2002} and
solid-state nuclear magnetic resonance (NMR) spectroscopy
\cite{collinsfoulkesbondklinowski1-1999,collinsduerklinowski321-2000}.

These kind of fullerene based materials have attracted a lot of attention
because of their unusual conducting and superconducting properties.
Solid \csixty\ is insulating but can be made 
conducting and even superconducting upon intercalation of alkali atoms
(e.g.~K, Rb) between the fullerene molecules
(see the reviews
\cite{hebard45-1992,gunnarson69-1997,gunnarsonkochmartin-1998}).
The alkali atoms transfer their valence electrons into the conduction bands of
the \csixty\ subsystem and the materials become metallic.

A noticeable enhancement of the superconducting critical temperature,
$T_c$,
of \csixty\ was expected when electrons are removed (hole doping)
because of the larger density of states of the valence bands.
Doping of $\sim 3$ holes per \csixty\ molecule and corresponding metalization 
was achieved with the use of field-effect doping technique
and has resulted in $T_c = 52$~K \cite{schoenklocbatlogg408-2000}.
Such doping has also been realized recently
\cite{ramirez-june2002,service296-2002}.
The intercalation of \csixty\ with haloform molecules has led to an further 
increase of the critical temperature \cite{schoenklocbatlogg293-2001}.
Such an enhancement of $T_c$ upon intercalation of molecules into a
superconductor has been already observed earlier
\cite{meunierburgerdeutscherguyon26-1968} and was discussed theoretically
\cite{kresin49-1974}.
In this work we focus on the normal properties of haloform intercalated
\csixty.

\xray\ powder diffraction measurements at various temperatures
\cite{jansenwaidmann14-1995,
 dinnebiergunnarssonbrummkochhuqstephensjansen-2002}
have shown that above $\sim$~200~K the crystal structure of haloform 
intercalated \csixty\ is hexagonal and transforms to a
triclinic structure below $T \sim 150$~K.
The stoichiometric composition of these materials is \csixtx\ (X=Cl,Br).
Solid state NMR measurements on \csixbr\ have shown that
above 193~K the \csixty\ molecules rotate isotropically
\cite{collinsduerklinowski321-2000}, similar to that found in pure 
face-centered cubic (fcc) \csixty.
Whereas below this temperature a rotation barrier of 6~\kjmol\ was
evaluated .
Also the bromoform molecules are found to be motionally active down to
218~K \cite{collinsfoulkesbondklinowski1-1999}.

Dinnebier \etal\
\cite{dinnebiergunnarssonbrummkochhuqstephensjansen-2002}
calculated the electronic structure of these materials using
a two-dimensional periodic tight-binding formalism where
only one specific lattice plane was considered.
They omitted the haloform molecules completely and took only the valence 
electrons of the five-fold degenerate $h_u$ molecular
orbital (MO) of each \csixty\ explicitely into account.
The main conclusion of their calculations is that the electronic density of 
states (DOS) at the Fermi energy for
$\sim 3$ holes per \csixty\ molecule satisfies following sequence:
DOS(\csixbr) $\approx$ DOS(\csixcl) $<$ DOS(\csixty).
Hence, the observed enhancement of $T_c$ cannot simply be explained by an 
increase of the DOS at the Fermi energy when \csixty\ is intercalated with 
haloform molecules and \clform\ is replaced by \brform\ ({\it vide supra}).
Given these observations and results, it is of interest to study
the electronic properties of these materials in more detail.

We have performed all-electron density functional (DF) calculations 
on several crystal structures of bulk \csixtx\ (X=Cl,Br) to predict
their electronic structures and to examine the effect of introduced
charge carriers (doping with electrons as well as with holes).
Employed are methods that rely on the generalized gradient approximation (GGA)
and the spin-unrestricted Kohn-Sham approach.
Different charge doping levels are simulated simply by removing or adding
electrons from the systems.
Unlike previous work \cite{dinnebiergunnarssonbrummkochhuqstephensjansen-2002}
our DF calculations include all electrons of the system into the quantum
treatment {\it and} the haloform molecules are taken into account
explicitly.
Particular attention is paid to the interaction of the
haloform molecules with the electronic structure of the \csixty\
subsystem. 
DF calculations have also been performed on an fcc and
on a hypothetical hexagonal lattice of \csixty\ molecules.
The latter system is formed by artificially removing the haloform molecules
from \csixtx.
These calculations allow to examine changes in the electronic structure
of the \csixty\ subsystem 
(i) upon intercalation and
(ii) due to electronic contributions of the haloform molecules separately.
Note, that all structures and electronic properties are obtained in the
absence of an applied electric field.

The structure of the article is as follows: 
Sec.~\ref{sec:xray-strcts} summarizes the experimental determination of the 
crystal structure of \csixtx\ (X=Cl,Br) Sec.~\ref{sec:comp-methods} 
gives details about the computational method. 
In Sec.~\ref{subsec:calc-structs} and Sec.~\ref{subsec:electr-structs} we 
discuss the calculated crystal and electronic structures of the un-doped 
systems, respectively.
Sec~.\ref{subsec:doping} then discusses the changes of the electronic
structure due to hole- and electron-doping of \csixtx.
Finally, a concluding discussion is found in Sec.~\ref{sec:concl}.

\section{crystal structures from \xray\ diffraction measurements}
\label{sec:xray-strcts}

\begin{figure*}
\begin{picture}(1780,1750)
%
\put(0,0){
\epsfxsize=178mm
\epsffile{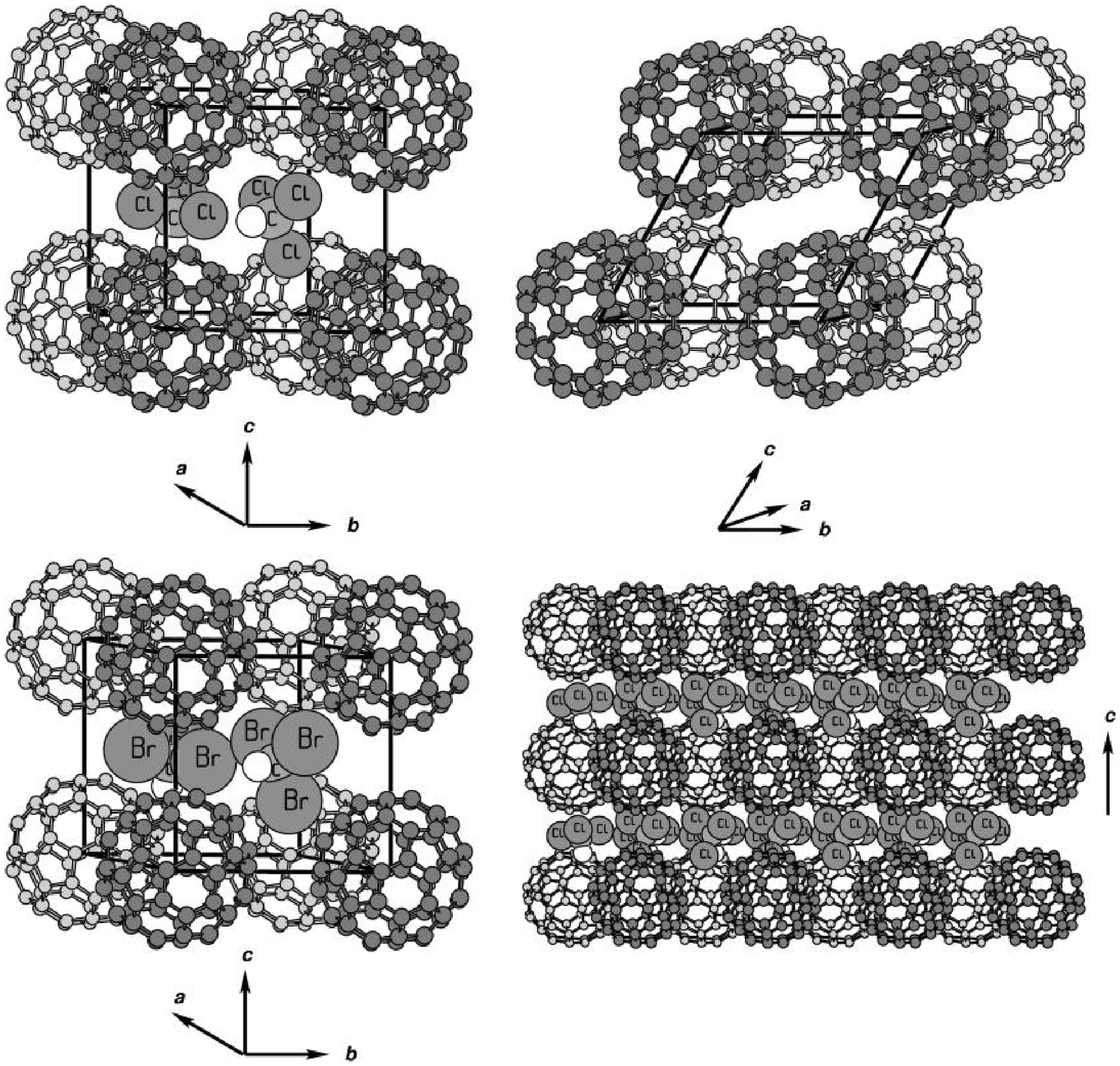}
}
\put(650,980){\mbox{\Large\bf (a)}}
\put(650,100){\mbox{\Large\bf (b)}}
\put(1600,980){\mbox{\Large\bf (c)}}
\put(1600,100){\mbox{\Large\bf (d)}}
\end{picture}
\caption{
Calculated configurations of (a) \clform\ and (b) \brform\ molecules
intercalated in hexagonal lattices of \csixty\ molecules.
Both crystal structures have P1 symmetry and are based on the
\xray\ powder diffraction analysis of Jansen and Waidmann
\cite{jansenwaidmann14-1995}.
(c) The molecules in pure \csixty\ form an fcc lattice and the crystal
structure has \fmthree\ symmetry.
Depicted is the primitive unit cell of fcc \csixty.
(d) The structures of hex.~\csixcl\ and hex.~\csixbr\ can be
considered as a sequence of alternating layers of \csixty\ molecules and
haloform molecules perpendicular to the $c$ vector.
}
\label{fig:calc-structs}
\end{figure*}
\begin{table}
\caption{
Crystal structures of pure \csixty\ and
\csixtx\ (X=Cl,Br) reported in the literature
and used in this work (unit cell lengths in {\AA}, angles in degrees).
Note, for \csixty\ the lattice parameters of the primitive cell is given.
}
\label{tab:lattice-param}
\begin{ruledtabular}
\begin{tabular}{lrrrrr}
 & \multicolumn{1}{l}{pure \csixty} & 
   \multicolumn{4}{l}{\csixtx\ (X=Cl,Br)} \\
 \cline{2-2} \cline{3-6}
temp. range [K] & \multicolumn{1}{l}{170} &
                  \multicolumn{2}{l}{$\geq$ 200} &
                  \multicolumn{2}{l}{$\leq$ 150} \\
Bravais lattice & \multicolumn{1}{l}{fcc} & 
                  \multicolumn{2}{l}{hexagonal} & 
                  \multicolumn{2}{l}{triclinic} \\
Space groups & \multicolumn{1}{l}{\fmthree} & 
               \multicolumn{2}{l}{P$\bar{3}$, P$\bar{1}$, P1} & 
               \multicolumn{2}{l}{P$\bar{1}$} \\
acronym & & \multicolumn{2}{l}{\bf HT~\tbn{a}} &
            \multicolumn{2}{l}{\bf LT~\tbn{b}} \\
Ref. & \multicolumn{1}{l}{\cite{david46-1995}} & 
               \multicolumn{2}{l}{\cite{jansenwaidmann14-1995}} & 
               \multicolumn{2}{l}
 {\cite{dinnebiergunnarssonbrummkochhuqstephensjansen-2002}} \\
 & \\
\multicolumn{6}{l}{\it lattice parameters} \\
 & & X=Cl & X=Br & X=Cl & X=Br \\
 \cline{3-4} \cline{5-6}
$a$      & 9.949 & 10.08 & 10.212 &  9.8361 &  9.8982 \\
$b$      & 9.949 & 10.08 & 10.212 & 10.0906 & 10.3386 \\
$c$      & 9.949 & 10.11 & 10.209 &  9.8179 &  9.8993 \\
$\alpha$ & 60.0  & 60.0  & 60.0   & 101.363 & 100.951 \\
$\beta$  & 60.0  & 60.0  & 60.0   & 116.457 & 115.920 \\
$\gamma$ & 60.0  & 120.0 & 120.0  &  79.783 &  78.202 \\
\end{tabular}
\end{ruledtabular}

\tbn{a}~High temperature structure.
\tbn{b}~Low temperature structure.
\hfil
\end{table}

The first crystal structures of \clform- and \brform- intercalated
\csixty\ reported in the literature were published by Jansen and Waidmann 
\cite{jansenwaidmann14-1995}.
An \xray\ powder diffraction analysis at room
temperature yielded for both compounds a primitive hexagonal lattice and an 
average symmetry of P6/mmm.
However, due to orientational disorder of the molecules the atomic coordinates
could not be resolved.
Throughout this paper this structure is labeled with HT (high temperature)
and the observed lattice parameters are summarized in 
Table~\ref{tab:lattice-param}.

Collins \etal\ examined the structure of \csixbr\ at room temperature
using \xray\ powder diffraction as well and found a hexagonal lattice with
unit cell parameters consistent with those for the HT structure of
Table~\ref{tab:lattice-param}.
They calculated the atomic positions using molecular modeling techniques
by placing a \csixty\ molecule at the origin of the hexagonal unit cell and
inserting the \brform\ molecules into the trigonal prismatic voids
at (1/3,2/3,1/2) and (2/3,1/3,1/2) [Fig.~\ref{fig:calc-structs}(b)].
Based on the modeled crystal structure they calculated an \xray\ diffraction 
pattern almost identical to the observed one.
The crystal structures can be considered as a sequence of alternating layers 
of \csixty\ and haloform molecules.
As an example, Fig.~\ref{fig:calc-structs}(d) shows the layer structure
of \csixcl.

Very recently, Dinnebier \etal\
\cite{dinnebiergunnarssonbrummkochhuqstephensjansen-2002}
performed high resolution \xray\ powder diffraction analyses at various
temperatures and determined the crystal lattices of different phases of
\csixtx\ (X=Cl,Br).
In agreement with Jansen and Waidmann and Collins \etal, the room 
temperature crystal lattice obtained is hexagonal and has P6/mmm space group 
symmetry.
At $T \approx 200$~K the haloform intercalated fullerenes undergo a 
first-order phase transition towards a monoclinic phase of C2/m symmetry which
has the double volume of the room temperature phase.
The alternating layers of haloform and \csixty\ molecules are shifted
with respect to each other such that the molecules are
displaced from their equilibrium positions in the room temperature phase.
Below $\sim$150~K a second first-order phase transition occurs and the crystal 
structures of \csixtx\ are triclinic and have \pbone\ symmetry
\cite{dinnebiergunnarssonbrummkochhuqstephensjansen-2002}.
This low temperature (LT) phase has similar cell dimensions as the hexagonal
room temperature structure (Table~\ref{tab:lattice-param}).
Throughout this paper this structure is labeled with LT.

A comparison of the hexagonal unit cells of \csixtx\ (X=Cl,Br)
[Fig.~\ref{fig:calc-structs}(a) and \ref{fig:calc-structs}(b)]
with the primitive unit cell of pure fcc \csixty\ 
[Fig.~\ref{fig:calc-structs}(c)] emphasizes the
structural changes caused by the intercalation with haloform molecules.
The cubic closed-packed (ccp) lattice of \csixty\ molecules transforms 
to a hexagonal close-packed lattice of \csixty\ molecules.
As a result of the intercalation the length of the primitive cell is 
increased by 1 to 4~\%.

\section{Computational method}
\label{sec:comp-methods}

All calculations are performed with the DMol$^3$ approach
\cite{delley921-1990,delley11318-2000}.
The calculations make use of the gradient corrected density functional of Becke
\cite{becke987-1993} for the exchange contribution, combined with the
Perdew Wang 1991 functional \cite{perdewwang4523-1992} for the
correlation contribution.
All electrons are described explicitely with a double numeric basis set 
augmented with polarization functions (DNP): 
2s1p for H, 3s2p1d for C, 4s4p2d for Cl and 5s4s2d for Br.
The cutoff radius to generate the atomic basis functions is set to
$7.0$~bohr (1 bohr~$ = 5.29177 \times 10^{-11}$~m).
The electron densities and the energies are calculated self consistently,
latter within an accuracy of 10$^{-5}$ hartree
(1 hartree$ = 2625.5$~\kjmol).
The numerical integration is performed on medium sized meshes.
For the Brillouin zone integration irreducible $\bdk$ vector meshes of 
different sizes are used \cite{delley11318-2000}.
Reliable energies and atomic positions are obtained with a
$\bdk$ vector mesh of $(n_x, n_y, n_z) = (4,4,2)$,
where $n_i$ denotes the increments along each of the primitive reciprocal 
lattice vectors, $\bdG_i$ (Fig.~\ref{fig:hex-bz}) \cite{delley11318-2000}.
This $\bdk$ vector mesh amounts to 20 symmetry-inequivalent sampling points.
The calculations of the DOS required a finer $\bdk$ vector mesh of
$(n_x, n_y, n_z) = (8,8,4)$, i.e.~132 irreducible $\bdk$ points.
Calculations on fcc \csixty\, on the other hand, require only
35 symmetry-unique sampling points, $(n_x, n_y, n_z) = (8,8,8)$.
Using the self-consistent electronic densities the electronic band structures 
of fcc \csixty\ and \csixtx\ (X=Cl,Br) are calculated at 59 and 
153 reciprocal lattice vectors, respectively, at the surface of the 
irreducible wedge of the appropriate Brillouin zones.
%

Doping with holes (electrons) is simulated by simply removing (adding)
an appropriate number electrons from the unit cell.
For instance, to simulate a doping of three holes per \csixty\
molecule, three electrons are removed and the total charge of the unit cell
is denoted $Q = +3$.
The resulting positive or negative charge of the system is compensated by 
a background counter charge (jellium approach).
Considered are doping levels of $Q = 0$, $Q = \pm 1$ and $Q = \pm 3$.
The charging procedure forms systems with open electron shells.
Therefore, all calculations on the charged systems are performed within the
unrestricted Kohn-Sham formalism to account for possible spin polarization
effects.

The crystal structures introduced in the next section for which the
electronic properties are determined, are optimized by keeping the cell
parameters fixed and allowing the atoms to move.
This concerns neutral as well as charged systems.
The atomic positions are optimized employing delocalized internal 
coordinates \cite{andzelmkingsmithfitzgerald335-2001}.
The convergence criteria of the optimization procedures are
$10^{-5}$~hartree for the energy, $3\times10^{-4}$~hartree/bohr for the
maximum energy gradient and $3\times10^{-3}$~bohr for the
maximum atomic displacement.

\section{Results and discussion}
\label{sec:res_disc}

\subsection{Optimized crystal structures}
\label{subsec:calc-structs}

\begin{table*}
\caption{
Relative energies between the different configurations (\dE, in \kjmol) and 
selected intermolecular atomic distances (in pm) of several calculated 
structures of neutral fcc \csixty\ and \csixtx (X=Cl,Br).
Summarized are the shortest distances and, if given, the corresponding 
average values in parenthesis.
Note, the distances of fcc \csixty\ are related to the primitive unit
cell [Fig.~\ref{fig:calc-structs}(c)]
}
\label{tab:calc-structs}
\begin{ruledtabular}
\begin{tabular}{lrrrrrrr}
 & \dE & \multicolumn{3}{l}{C$\cdots$C (\csixty)} &
 C$\cdots$X~\tbn{a} & X$\cdots$X~\tbn{b} & C$\cdots$H~\tbn{b} \\
 \cline{3-5}
 & & along~$\bda$, $\bdb$\hfill & along~$\bda+\bdb$ & along~$\bdc$ &
 \multicolumn{1}{c}{(\csixty)} & & \multicolumn{1}{c}{(\csixty)} \\
\hline
\multicolumn{1}{l}{fcc \csixty~\tbn{c}} \\
 \hfill \fmthree &   & 299, 301~\tbn{f} & 315~\tbn{f} & 342~\tbn{f} \\
\multicolumn{1}{l}{\csixcl} \\
 \hfill \poneht~\tbn{d}          &  0 &
  340, 302 & 340 & 354 & 301 (339) & 308 (348) & 262 (347) \\
 \hfill \pbthree~\tbn{d}    &  3 &
  306, 306 & 306 & 362 & 355       & 350       & 334 (359) \\
 \hfill \ponep~\tbn{d}      & 11 &
  304, 302 & 346 & 355 & 317 (338) & 298 (328) & 295 (345) \\
 \hfill \pboneht~\tbn{d} & 14 &
  303, 302 & 351 & 356 & 319 (342) & 305 (326) & 303 (357) \\
 \hfill \pbonelt~\tbn{e} & 28 &
  326, 324~\tbn{g} & 340~\tbn{g} & 361~\tbn{g} &
     339 (347) & $\cdots$  & 269 (335) \\
\multicolumn{1}{l}{\csixbr} \\
 \hfill \poneht~\tbn{d} &  0 &
  354, 335 & 343 & 360 & 300 (341) & 318 (342) & 251 (342) \\
 \hfill \pbonelt~\tbn{e} &  1 &
  332, 334~\tbn{g} & 354~\tbn{g} & 385~\tbn{g} &
     343 (349) & $\cdots$  & 286 (337) \\
\end{tabular}
\end{ruledtabular}

\tbn{a}~Distances shorter than 357 pm.
\tbn{b}~Distances shorter than 400 pm.
\tbn{c}~Crystal structure of David \cite{david46-1995}.
\tbn{d}~Crystal structure of Jansen and Waidmann
        \cite{jansenwaidmann14-1995}.
\tbn{e}~Crystal structure of Dinnebier \etal\
        \cite{dinnebiergunnarssonbrummkochhuqstephensjansen-2002}.
\tbn{f}~The $\bdc$ vector of the hexagonal unit cell corresponds to the
        $\bdb$ vector of the primitive fcc unit cell.
\tbn{g}~The $\bdc$ vector of the hexagonal unit cell corresponds to the
        $\bdb$ vector of the triclinic unit cell used in
        Ref.~\cite{dinnebiergunnarssonbrummkochhuqstephensjansen-2002}.
\hfil
\end{table*}

Calculations are performed on the hexagonal HT structures
of Jansen and Waidmann \cite{jansenwaidmann14-1995} as well as
on the triclinic LT structures of Dinnebier \etal\
\cite{dinnebiergunnarssonbrummkochhuqstephensjansen-2002}.
Because the atomic positions in the unit cells of the HT structures are 
unknown, several different atomic configurations are designed
as starting points for the structure optimizations.
Physically and chemically reasonable configurations can only be designed 
when the experimental determined space group symmetry P6/mmm is reduced.
%

One designed configuration of \csixcl\ has P$\bar{3}$ symmetry.
The two chloroform molecules in the unit cell are located in the trigonal 
prismatic voids at (1/3,2/3,1/2) and (2/3,1/3,1/2), respectively, and are
related by inversion symmetry.
One six-fold axis of the \csixty\ and the $C_3$-axes of the 
\clform\ molecules coincide with the $c$ axis of the hexagonal lattice.
The two C-H bonds point towards the $-\bdc$ and $+\bdc$ lattice vector, 
respectively.
Three other start configurations are derived from the \pbthree\ 
configuration by exchanging the hydrogen atoms of the \clform\ molecules 
with arbritrary chlorine atoms.
Two of these configurations have P1 symmetry and one has
\pbone\ symmetry.
Due to the exchange of H and Cl atoms the \clform\ molecules are
differently oriented and probably induce also an orientational change of
the \csixty\ molecules.
Throughout this article all configurations are named by their symmetry labels,
i.e.~the configurations formed are termed \poneht, \ponep, \pbthree\
and \pboneht.
The most stable configuration of \csixcl, i.e.~the one with the lowest
calculated energy, is used to form the starting configuration for the
structure optimizations of \csixbr.

For the calculations on pure fcc \csixty\ the crystal structure of
David is employed \cite{david46-1995}.
The space group of this structure is \fmthree, i.e.~all
molecules have the same orientation.
Its primitive unit cell contains one molecule [Fig.~\ref{fig:calc-structs}(c)]
and has a length of $a = 9.949$~{\AA} (Table~\ref{tab:lattice-param}).

It is worth noting that both \clform\ and \brform\ have permanent electric
dipole moments of $\mu \sim 1$~D \cite{handbookchemphys-1999} and
the \csixty\ molecules are highly polarizable 
($\alpha \sim 80$~{\AA}$^3$ \cite{ruizbretongomezllorente1143-2001}).
The charge distribution within the haloform molecules is such that the
halogen atoms are charged partially negatively and the hydrogen atoms are
charged partially positively.
Hence, the stability of \csixtx\ (X=Cl,Br) is 
presumably mainly due to dipole-induced dipole interactions rather than by 
van-der-Waals interactions between the molecules.

Table~\ref{tab:calc-structs} summarizes the relative energies, \dE, between
the optimized crystal structures of neutral \csixcl\ and \csixbr.
All energies are with respect to the most stable configurations which have
relative energies of \dE = 0~\kjmol.
The most stable structures of \csixcl\ are the \poneht\ 
configurations [Fig.~\ref{fig:calc-structs}(a)].
The energy difference to the next stable configuration (\pbthree) is 
only 3~\kjmol.
Although the LT configurations correspond to temperatures below $\sim$150~K and 
the HT structures to room temperature, the calculations on \csixcl\
yields a lower energy for the \poneht\ configuration
than for the \pbonelt\ configuration.
The energy difference is \dE$ = 28$~\kjmol.
In the case of \csixbr\ both structures, \poneht\ and
\pbonelt, have nearly identical energies.

These energy considerations are made for zero temperature and, therefore,
do not include any entropic effects.
The energies shall give only a clue about the shape of the potential energy 
surface of \csixtx\ (X=Cl,Br).
The small energy differences between the configurations give rise to a shallow
potential energy surface provided that the energy barriers of
molecular re-orientations are of the same order as the relative energies.
Solid-state NMR measurements predicted for the re-orientations of the
\csixty\ molecules in \csixbr\ activation energies of 6~\kjmol
\cite{collinsduerklinowski321-2000}.
Nevertheless, in order to give a more precise answer of the relative stability 
between the configurations \csixcl\ and \csixbr\, at least the zero-point 
vibrational energy and the vibrational partition function need to be 
calculated.

All optimized structures of \csixtx\ (X=Cl,Br), obtained from the
HT and LT data have almost identical intramolecular atomic distances
(in pm):
\dist{C-C} = 145, \dist{C=C} = 140, \dist{C-H} = 109, \dist{C-Cl} = 177--179,
\dist{C-Br} = 194--196.
These distances are equivalent to the appropriate distances of free
molecules in the gas phase.

The main differences between the calculated structures can be found 
when comparing the intermolecular distances.
The differently oriented haloform molecules in the start configuration of 
the optimization procedure induce changes in the orientation of the 
fullerene molecules.
We note that for both chloroform and bromoform intercalation
the high resolution \xray\ measurements
\cite{dinnebiergunnarssonbrummkochhuqstephensjansen-2002} and
the calculated LT structures (this work) are almost identical.
The largest differences between these structures are in the intermolecular
distances and are smaller than 5~pm.
This means, the optimization of the atomic positions of the LT structures
with fixed lattice parameters does not cause any drastical structural changes.

Table~\ref{tab:calc-structs} summarizes selected
intermolecular distances of the optimized crystal structures of neutral 
fcc \csixty\ and of neutral \csixtx\ (X=Cl,Br).
While the expansion of the lattice of \csixty\ molecules caused by
the intercalation is accompanied by an increase of the unit cell length of
only 1 to 4~\% the changes of the shortest distances between \csixty\
molecules vary between 1 and 18~\%.
This indicates that there are changes in the orientation of the molecules.
Due to the larger effective volume of \brform\ the shortest interatomic
distances between neighboring \csixty\ molecules are on the average larger 
for \csixbr\ than for \csixcl\ and fcc \csixty.
But the largest changes are found for the shortest distance between C atoms of 
nearest neighboring \csixty\ molecules along the cell
vector $\bda$, \dist{C$\cdots$C}$_{\bda}$, of the \poneht\ structures,
which are 41 (X=Cl) and 53~pm (X=Br).
Note, that the distances of fcc \csixty\ are related to the primitive unit
cell [Fig.~\ref{fig:calc-structs}(c)].

All the shortest C$\cdots$C distances along the $\bdc$ vector,
\dist{C$\cdots$C}$_{\bdc}$, of the HT as well as the LT structures
are larger than the shortest C$\cdots$C distances in the 
(0001) plane, typical for layered system structures.
By far the largest \dist{C$\cdots$C}$_{\bdc}$ value of 385~pm is for the
\pbonelt\ configuration of \csixbr.

Both in fcc \csixty\ and the HT structures of \csixcl\ there is at 
least one value of \dist{C$\cdots$C}$_{\bda}$ or \dist{C$\cdots$C}$_{\bdb}$
which is around 300~pm.
These distances are about 40~pm smaller than the sum of the van-der-Waals 
radii of two carbon atoms ($R_{vdW}$(C)$ = 170$~pm \cite{webelements}).
This is evidence for an overlap between $\pi$ orbitals of neighboring 
\csixty\ molecules.
While the \pbonelt\ and most of the HT structures of \csixcl\ have
almost identical values for \dist{C$\cdots$C}$_{\bda}$ and
\dist{C$\cdots$C}$_{\bdb}$ only \poneht\ \csixcl\ has two different
values, where \dist{C$\cdots$C}$_{\bda}$ is about 40~pm larger than
\dist{C$\cdots$C}$_{\bdb}$.
Because of symmetry restrictions for \pbthree\ \csixcl,
all the shortest intermolecular C$\cdots$C distances in the (0001) plane, 
\dist{C$\cdots$C}$_{\bda}$, \dist{C$\cdots$C}$_{\bdb}$ and 
\dist{C$\cdots$C}$_{\bda + \bdb}$, are equivalent.

On the other hand the LT configuration of \csixcl\ (\pbonelt) does not 
have any interatomic distance between neighboring \csixty\ molecules 
which is shorter than 324~pm.
But the values of \dist{C$\cdots$C}$_{\bda}$ and \dist{C$\cdots$C}$_{\bdb}$
are also almost identical, similar to fcc \csixty.

%
%

In both structures of \csixbr\ considered only \dist{C$\cdots$C}$_{\bdb}$
are shorter than 340~pm, the sum of the van-der-Waals radii of two carbon
atoms.
Similar to \csixcl\ the \poneht\ structure of \csixbr\ has
two different values for \dist{C$\cdots$C}$_{\bda}$ and
\dist{C$\cdots$C}$_{\bdb}$ while for \pbonelt\ these distances are
almost identical.

For all configurations both the average values and the smallest values of
the distances between the halogen atoms X (X=Cl,Br) and the atoms of the
\csixty\ molecules are smaller than the sum of the corresponding van-der-Waals
radii ($R_{vdW}$(Cl)$ = 175$~pm, $R_{vdW}$(Br)$ = 185$~pm \cite{webelements}).
The \poneht\ configurations of \csixcl\ and \csixbr\ have the 
smallest value of \dist{C$\cdots$X}$ \approx 300$~pm.
This indicates a significant intermolecular overlap between orbitals of
\csixty\ and CHX$_3$.

Due to their partial negative charge the halogen atoms of neighboring haloform
molecules should avoid each other.
Therefore, it is peculiar that the majority of the configurations have
halogen atoms X that are closer than twice of $R_{vdW}$(X).
Only the \pbonelt\ structures do not have any X$\cdots$X distance
smaller than 400~pm.

All configurations have similar average values of the non-bonding C$\cdots$H
distances.
The \poneht\ structures have the shortest non-bonding C$\cdots$H distances.
Apparently, the partially positively charged H atoms are attracted by 
the negatively charged $\pi$ electron system of the \csixty\ molecules.

The electronic structure calculations of the following sections are performed
representatively for the most stable (\poneht) and most unstable
(\pbonelt) configurations as determined in this section.
It is assumed that the other structures have similar properties which can
be derived or approximated from the properties of the \poneht\ and 
\pbonelt\ configurations.

\subsection{Electronic structures}
\label{subsec:electr-structs}

\subsubsection{Band structures and DOS}
\label{subsubsec:bands}

%
%
%
\begin{figure}
\begin{picture}(860,650)
%
\put(0,0){
\epsfxsize=86mm
\epsffile{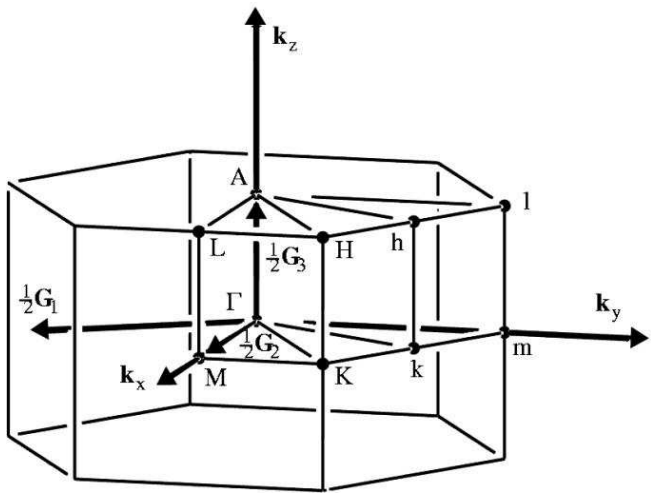}
}
\end{picture}
\caption{
Hexagonal Brillouin zone with reduced symmetry. \hfill
}
\label{fig:hex-bz}
\end{figure}
\begin{figure*}
\begin{picture}(1780,830)
%
\put(0,-10){
\epsfxsize=83mm
\epsffile{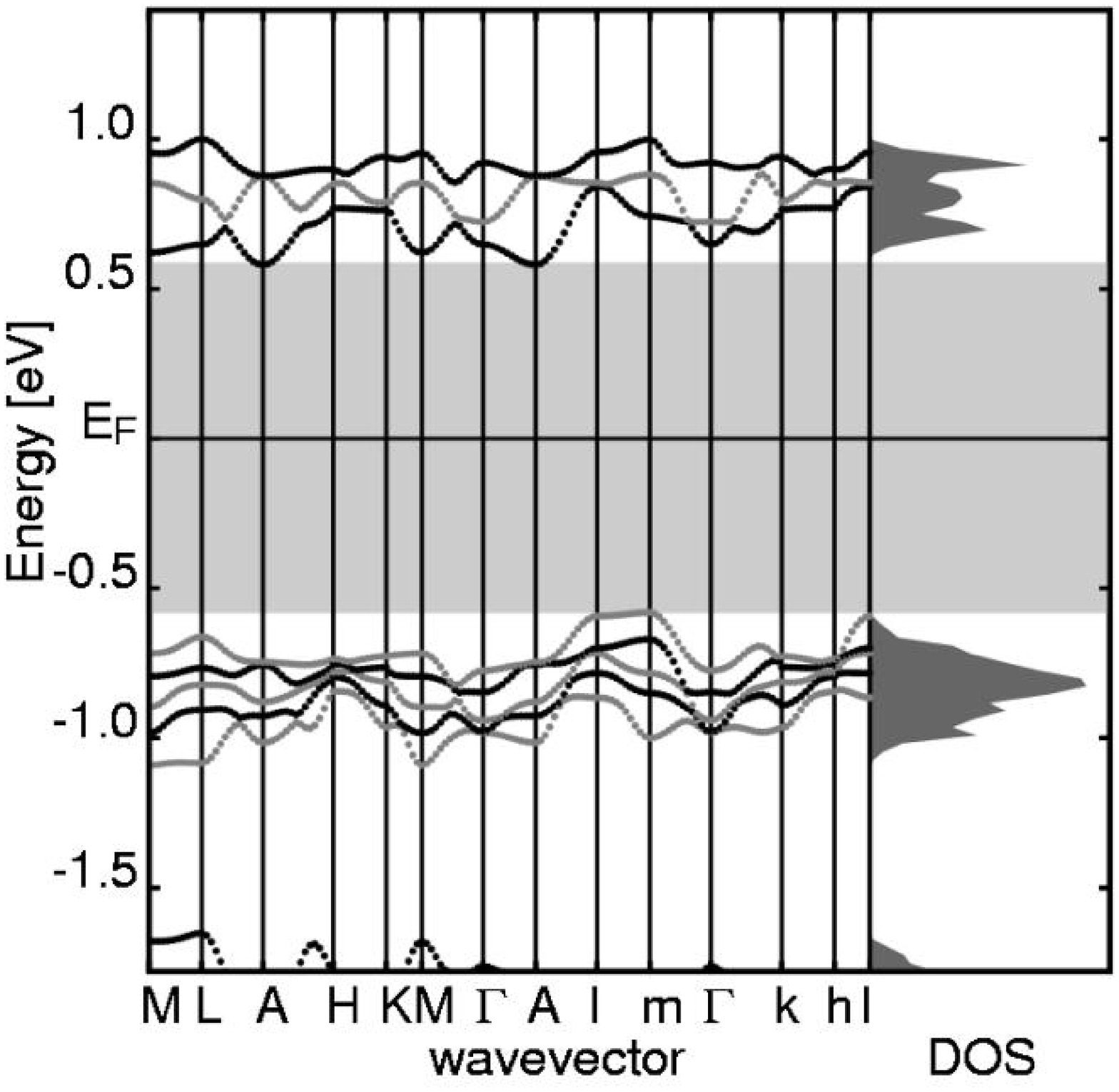}
}
\put(450,820){\bf \csixcl}
%
%
\put(900,-10){
\epsfxsize=83mm
\epsffile{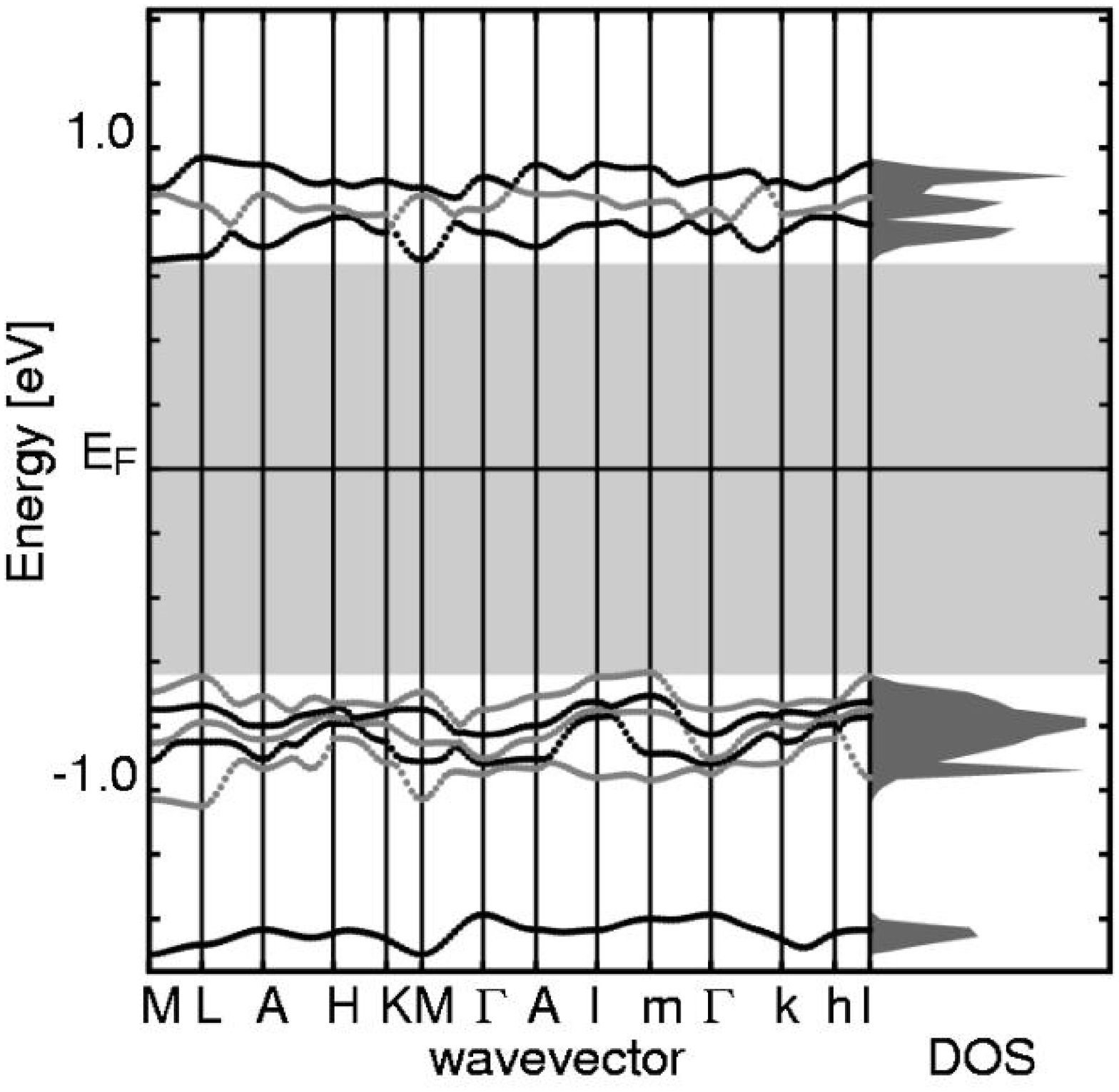}
}
\put(1380,820){\bf \csixbr}
\end{picture}
\caption{
One-electron band structures and total density of states (DOS) of neutral 
\poneht\ \csixtx\ (X=Cl,Br).
All energies are with respect to the Fermi energy, \Efermi.
The band structures depict the three conduction bands, the five valence
bands and the next state below the valence bands.
For clarity, the bands are gray and black in alternating order.
}
\label{fig:bands}
\end{figure*}
%
%
%

%
%
%
\begin{figure}
\begin{picture}(860,1730)
%
\put(0,-30){
\epsfxsize=83mm
\epsffile{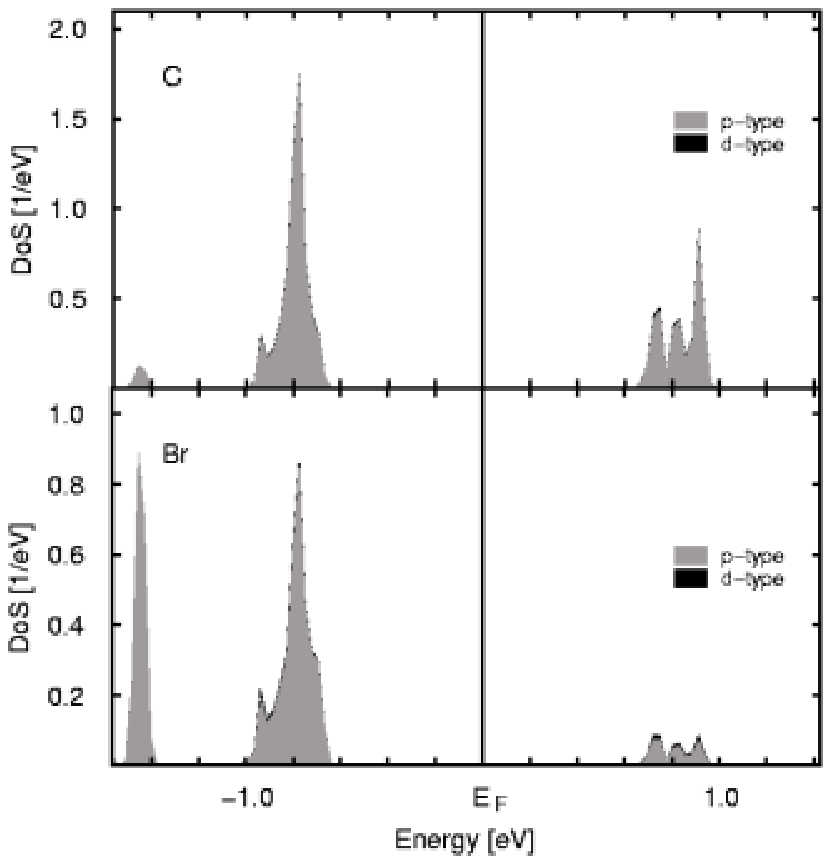}
}
\put(580,850){\bf \csixbr}
%
\put(-10,870){
\epsfxsize=83mm
\epsffile{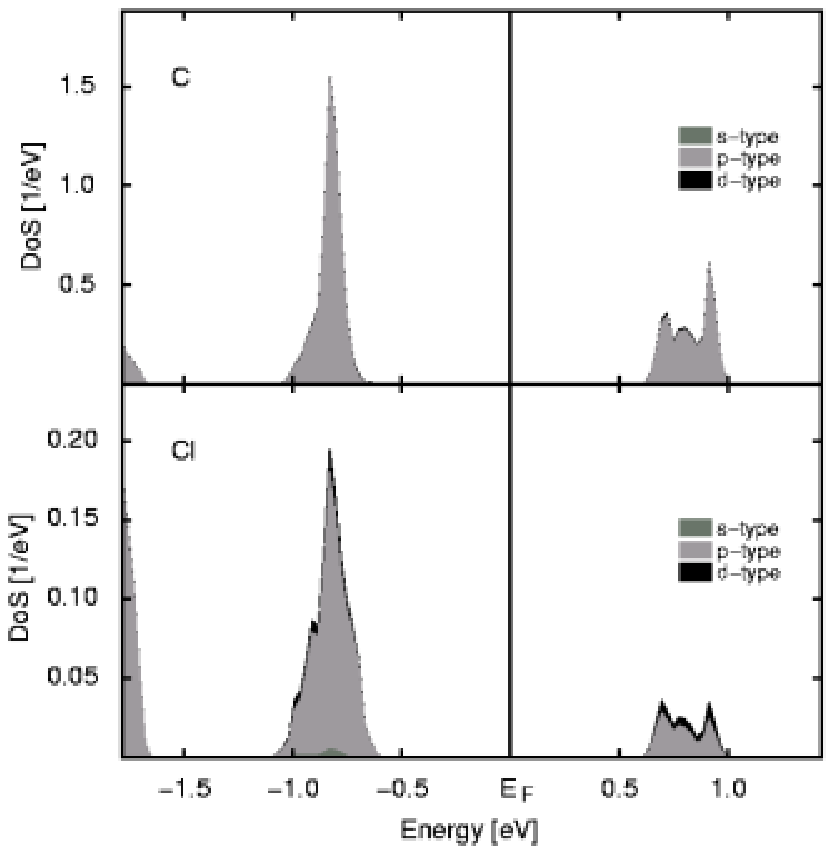}
}
\put(580,1730){\bf \csixcl}
%
\end{picture}
\caption{
Partial DOS of the bands depicted in Fig.~\ref{fig:bands}.
All energies are with respect to the Fermi energy, \Efermi.
The partial DOS show the contributions of the $s$-,
$p$- and $d$-type of orbitals of
specific atoms of the \csixty\ molecule, C, and of the haloform
molecules, Cl and Br, to the total DOS of \poneht\ \csixtx\
(X=Cl,Br).
Note the different ranges of the ordinates.
}
\label{fig:pdos}
\end{figure}
\begin{table}[b]
\caption{
Comparison of width and smallest indirect gaps for the
valence and conduction bands, VB and CB, respectively (in eV) calculated
for fcc \csixty\ and neutral \csixtx\ (X=Cl,Br).
}
\label{tab:width-gaps}
\begin{ruledtabular}
\begin{tabular}{lrrr}
 & \multicolumn{1}{l}{band gap} & \multicolumn{2}{l}{band width} \\
 \cline{2-2} \cline{3-4}
 & VB -- CB & VB & CB \\
\hline
fcc \csixty~\tbn{a} & 1.0 & 0.7 & 0.6 \\
\csixcl \\
\hspace{3mm} \poneht~\tbn{b}                     & 1.2 & 0.5 & 0.4 \\
\hspace{3mm} \poneht, without \clform~\tbn{b),c} & 1.2 & 0.5 & 0.4 \\
\hspace{3mm} \pbonelt~\tbn{d}                    & 1.1 & 0.6 & 0.4 \\
\csixbr \\
\hspace{3mm} \poneht~\tbn{b}  & 1.3 & 0.4 & 0.3 \\
\hspace{3mm} \pbonelt~\tbn{d} & 1.2 & 0.5 & 0.3 \\
\end{tabular}
\end{ruledtabular}
\tbn{a}~Crystal structure of David \cite{david46-1995}.
\tbn{b}~Crystal structure of Jansen and Waidmann
        \cite{jansenwaidmann14-1995}.
\tbn{c}~The two \clform\ molecules are simply deleted from the unit cell
        of the \poneht\ structure and the remaining \csixty\ molecule
        is re-optimized.
\tbn{d}~Crystal structure of Dinnebier \etal\
        \cite{dinnebiergunnarssonbrummkochhuqstephensjansen-2002}.
\hfil
\end{table}

Fig.~\ref{fig:bands} shows the valence and conduction bands of neutral
\poneht\ \csixtx\ (X=Cl,Br).
The corresponding Brillouin zone is depicted in Fig.~\ref{fig:hex-bz}.
The systems have indirect band gaps of 1.2 (X=Cl) and 1.3~eV (X=Br).
Since neutral \csixtx\ has empty conduction bands the systems are non-metallic.
Their valence bands (VB) and conduction bands (CB) consist of
five and three narrow bands, respectively, in accordance with the respective 
degeneracy of the constitutive molecular orbitals.
The total DOS of the CB clearly show three nearly separated
peaks, each belonging to one of the three bands.
None of the VB and CB are degenerate at any $\bdk$ vector of the
band structure.
The largest band dispersions are along $k_x$ and $k_y$.
Table~\ref{tab:width-gaps} collects the width of the VB and CB and the
smallest indirect gaps between VB and CB.
The widths of the VB and CB of \poneht\ \csixbr\ are 0.4 and 
0.3~eV, respectively.
The VB of \poneht\ \csixcl\ is 0.1~eV broader, because the fullerene
molecules are closer to each other (see Tab.~\ref{tab:calc-structs}) and, 
hence, the overlap of the $\pi$ orbitals of neighboring molecules is larger.
Note, that the band structure of neutral \csixbr\ has a distinguishing feature
namely that a single narrow band appears just 0.4~eV below the VB.
In the case of \csixcl\ the electronic states below the VB are rather
a group of bands.
The gap between the latter and the VB is about 0.5~eV.

The LT structures have similar band structures as the \poneht\ structures
but with band gaps of 1.1 (\csixcl) and 1.2 (\csixbr).
For both, \csixcl\ and \csixbr, the width of the VB is 0.1~eV larger than 
that of the appropriate HT structure in agreement with the intermolecular
distances of Table~\ref{tab:calc-structs}.
The widths of the CB and the gaps between the VB and the upper 
edge of the next lower lying electronic states are identical for the HT and LT
structures.

For fcc \csixty\ the calculated band gap is only 1.0~eV and the
valence and conduction bands are $0.1$ to $0.3$~eV broader than the VB and CB 
of the HT and LT structures.
Both, the smaller band gap and the broader bands are due to smaller distances 
between the fullerene molecules as compared to \csixtx\ (X=Cl,Br).

In order to disentangle the contribution of the chloroform molecules to the 
electronic structure of the \csixty\ subsystem in \csixcl,
calculations are also performed on a hypothetical hexagonal
lattice of \csixty\ molecules in the geometry of \poneht\ \csixcl.
This model system is designed by simply removing the two \clform\
molecules from the unit cell of \poneht\ \csixcl.
The shape and the positions of the valence and conduction bands almost do not
change.
The band gap of 1.2~eV and the widths of the VB and CB are identical to those
of \poneht\ \csixcl.
Hence, the chloroform as well as the bromoform molecules apparently
do not affect the band structure of the materials substantially.

One of the well-known drawbacks of DF calculations is the underestimation
of band gaps.
For instance, the experimental determined band gap of fcc \csixty\ is $2.3$~eV 
\cite{lofveenendaalkoopmansjonkmansawatzky6826-1992}
whereas the calculated gap is less than half of that ({\it vide supra}).
Hence, the band gaps presented in this article are only lower bounds.
The problem of too small band gaps in DF calculations can be
cured according to Slater transition-state methods
\cite{slater6-1972,liberman6211-1999}.

\subsubsection{Partial DOS}
\label{subsubsec:pdos}

Fig.~\ref{fig:pdos} shows the partial density of states (PDOS) for
specific atoms of \poneht\ \csixtx\ (X=Cl,Br).
The PDOS is the decomposition of the electron bands into contributions of
atomic orbitals (AO).
The atoms of \csixty\ and the haloform molecules chosen for the PDOS have 
the smallest intermolecular C$\cdots$X distances.
As expected, the VB and the CB of \poneht\ \csixcl\ and
\csixbr\ mainly consist of carbon $2p_{\pi}$ AO.
Surprisingly, also chlorine $3p$ and bromine $4p$ AO
contribute significantly to the PDOS of \poneht\ \csixcl\
and \csixbr, respectively.
However, the latter two contributions are smaller than that of the 
\csixty\ atoms.
The single electronic band just below the VB of \poneht\ \csixbr\ 
(Fig.~\ref{fig:bands}) mainly consists of bromine $4p$ AO.

The calculated PDOS for the LT structures leads to conclusions similar
to those for the HT structures.
However, the contributions of the $p$-type orbitals of the halogen atoms
are smaller and the contribution of fullerenes correspondingly larger,
because the LT structures have larger intermolecular
C$\cdots$X (X=Cl,Br) distances on average.

Consequently, the haloform molecules do not only have the function of inert
spacers to expand the \csixty\ lattice as inferred in
Section~\ref{subsubsec:bands} but might also support charge 
transfer and electron-phonon coupling.

\subsubsection{Charge distribution}
\label{subsubsec:charge-distr}

This section concerns only neutral \poneht\ \csixcl\ because 
neutral \poneht\ \csixbr\ and
neutral \pbonelt\ \csixtx\ (X=Cl,Br) have akin
charge density distributions.

\begin{figure}
\begin{picture}(860,1110)
%
\put(0,-20){
\epsfxsize=83mm
\epsffile{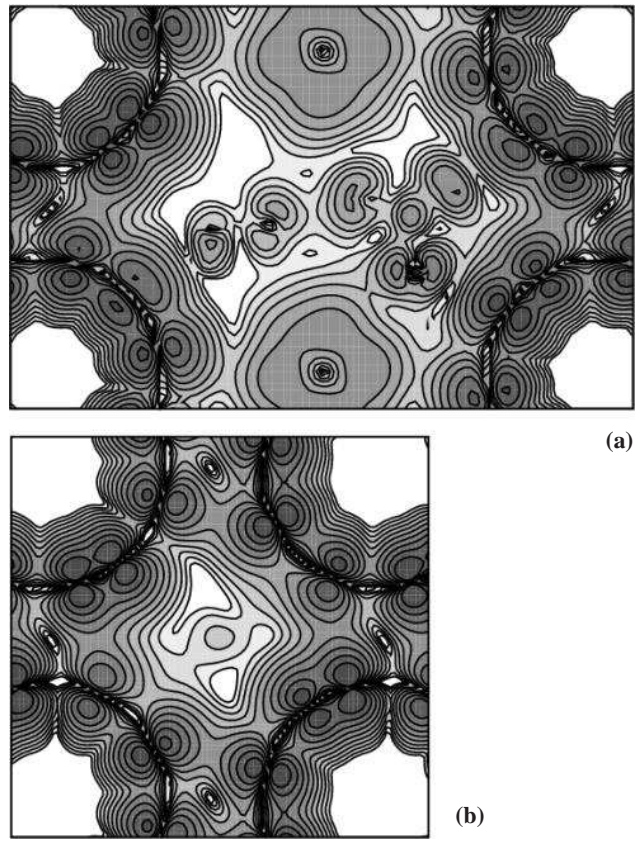}
}
\put(800,500){\mbox{\bf (a)}}
\put(600,0){\mbox{\bf (b)}}
\end{picture}
\caption{
Contour plots of the sum of the squares of the five highest occupied
crystal orbitals at the $\Gamma$ point of neutral \poneht\ \csixcl\
in the [(a)] ($\bar{1}100$) and [(b)] ($1100$) planes.
The iso-lines depict values from $10^{-5}$ to $5.12 \E{-3}$ each
increased by a
factor of two (1~a.u.~$ = 6.7483 \E{30}$~electrons~$\cdot$~m$^{-3}$).
The darker the gray the higher is the electron density.
}
\label{fig:homodens}
\end{figure}

Figs.~\ref{fig:homodens}(a) and \ref{fig:homodens}(b) show contour plots of the
valence charge density in the ($\bar{1}100$) and ($1100$) planes,
respectively.
The valence charge density is the sum of the squares of the five highest 
occupied crystal orbitals of neutral \csixcl. The integral over the
Brillouin zone is obtained using the mid-point approximation ($\Gamma$
point).
In the corner of the maps are parts of the \csixty\ molecules
represented by their $\pi$ electron systems.
Clearly visible are the carbon $2p_{\pi}$ AO.
In the center of Fig\ref{fig:homodens}(a) valence charge density that
belongs to the two \clform\ molecules can be seen.
Both density plots show considerable overlap of the $p_\pi$ orbitals of
adjacent \csixty\ molecules as well as overlap of the chlorine $p$ 
orbitals with the $\pi$ electron systems of the \csixty\ molecules.
The latter overlap is smaller than the overlap between the $\pi$ electron
systems, yet relevant.

\begin{figure*}
\begin{picture}(1780,480)
%
\put(0,-20){
\epsfxsize=178mm
\epsffile{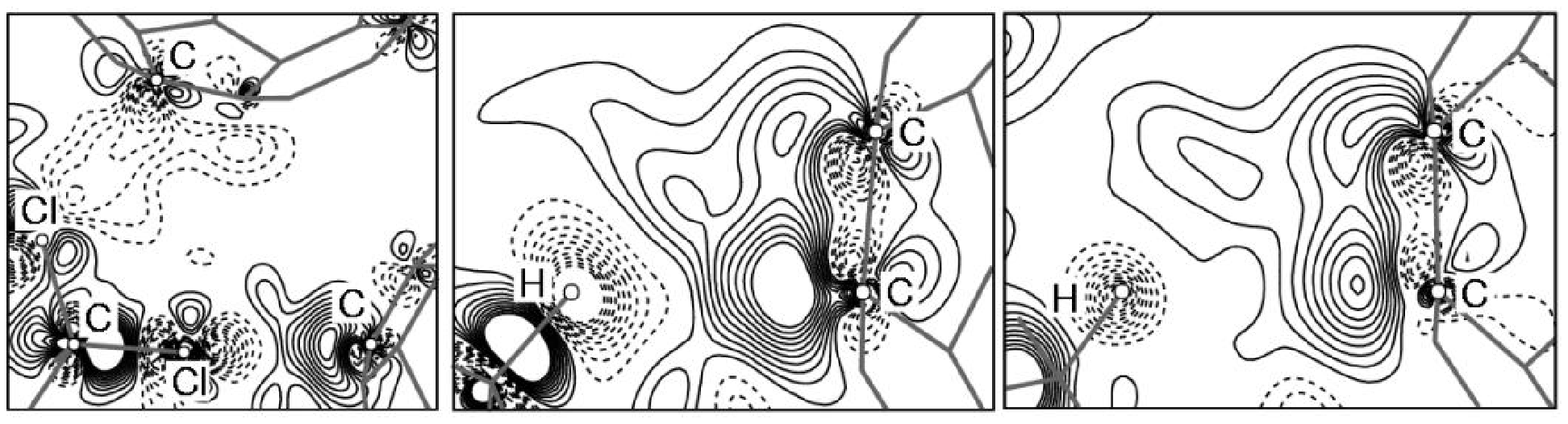}
}
\put(50,400){\bf (a)}
\put(580,400){\bf (b)}
\put(1200,400){\bf (c)}
\end{picture}
\caption{
Contour plots of the molecular deformation density, $\rho_{diff}$, plotted 
for relevant planes of \poneht\ \csixcl\
(see text for the definition of $\rho_{diff}$).
(a) shows the changes of the total density induced by Cl atoms,
and (b) and (c) induced by the two H atoms in the unit cell.
The atoms in and near the plane are marked by open circles.
The iso-lines represent values from $-10^{-4}$ to $10^{-4}$~a.u.~in steps of
$10^{-5}$~a.u.~.
Dashed lines are for negative and solid lines for positive values.
The light-gray solid lines sketch parts of the \clform\ and \csixty\
molecules.
}
\label{fig:deformdens}
\end{figure*}

The contour plots of Fig.~\ref{fig:deformdens} show the molecular deformation
density, $\rho_{diff}$, in particular planes.
The molecular deformation density is defined as the difference between the 
total electron density of \csixcl\ and the sum of the densities of the
subsystems \csixty\ and \clform.
Note, that both subsystems have the same geometry as in \csixcl.
The molecular deformation density discloses changes of the electron density of
the \csixty\ and chloroform molecules due to intercalation,
e.g.~polarization effects or charge transfer from one molecule to another.

Fig.~\ref{fig:deformdens}(a) shows the changes of the electron density
induced by chlorine atoms that have smallest interatomic distances to
the \csixty\ molecules.
One chlorine atom pulls electron density away from \csixty\ molecule and
the other chlorine atom pushes electron density towards a \csixty\
molecule.
Both, Fig.~\ref{fig:deformdens}(b) and Fig.~\ref{fig:deformdens}(c), depict
the polarization of carbon-carbon $\pi$ bonds due to the two partially 
positively charged hydrogen atoms of the chloroform molecules.
These hydrogen atoms have also the smallest interatomic distances to 
\csixty\ atoms.
The electron density surrounding the $\pi$-bonds is increased at the
expense of the density around the hydrogen atoms.

Note that the deformation density is of the order of $10^{-4}$ atomic units
and, hence, very small.
However, the contour plots of Fig.~\ref{fig:deformdens} clearly validate
the presumption of dipole-induced dipole interactions between the
haloform and the \csixty\ molecules.

\subsection{Changes of the electronic structures due to charging}
\label{subsec:doping}

\begin{figure*}
\begin{picture}(1660,1100)
%
%
\put(0,1080){\mbox{\csixcl}}
\put(0,1030){\mbox{\poneht}}
\put(450,1030){\mbox{\pbonelt}}
\put(900,1080){\mbox{\csixbr}}
\put(900,1030){\mbox{\poneht}}
\put(1350,1030){\mbox{\pbonelt}}
\put(-70,0){
\epsfxsize=178mm
\epsffile{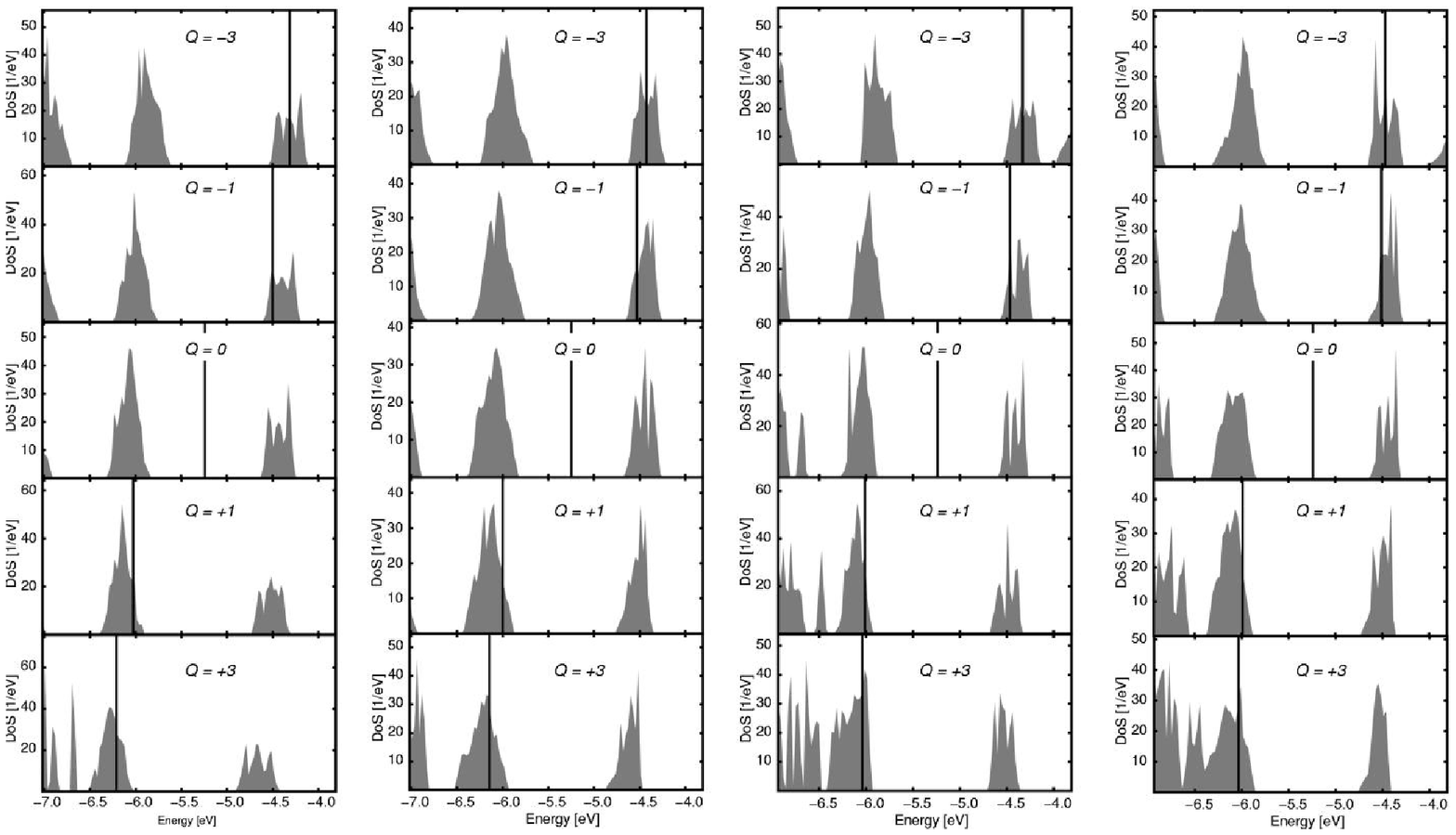}
}
\end{picture}
\caption{
Total density of states (DOS) around the Fermi energy, \Efermi, of the
neutral and charged \poneht\ and \pbonelt\ structures of
\csixtx\ (X=Cl,Br).
$Q$ is the total charge per unit cell.
The vertical solid lines represent \Efermi.
The corresponding total DOS at \Efermi\ are summarized in
Tab.~\ref{tab:charges-and-dos}.
}
\label{fig:charges-and-tdos}
\end{figure*}

In order to clarify how electron and hole doping affect the 
electronic structure and perhaps also the atomic configurations of 
haloform intercalated \csixty, calculations are performed on the charged
\poneht\ and \pbonelt\ structures of \csixtx\ (X=Cl,Br) as well as
on charged fcc \csixty.
The structural changes of the \poneht\ and \pbonelt\ configurations 
of \csixtx\ due to charging are found to be small.
The charging has no effect on the intramolecular distances.
For example Jahn-Teller distortion could not be found.
The changes in intermolecular distances of Table~\ref{tab:calc-structs} 
are of the order of 2~pm and are caused by small changes in the rotational
orientation of the haloform and \csixty\ molecules.
%
%
%

The Hirshfeld population analysis 
\cite{hirshfeld44-1977}
predicted that in each case 80~\% to 90~\% of the charge is located on the 
\csixty\ molecules and the remaining part on the haloform molecules.
%
%
%
%
%
%
%

Fig.~\ref{fig:charges-and-tdos} shows the total DOS of the five valence and
three conduction bands of neutral \csixtx\ (X=Cl,Br) and their
modifications upon hole and electron doping.
In the following the terms VB and the CB are related to the corresponding 
states of the neutral systems.
Overall the DOS of the VB and CB are narrow peaks with steep flanks.
However, for the same charge the shape of the DOS of the HT and LT structures
show differences.
For almost all charges the CB has three distinct peaks which indicates that 
it consists of three narrow bands.
The widths of the CB are in the range of $0.3$--$0.5$~eV without any clear 
trend with respect to charging.
The width of the DOS of the five valence bands is between $0.4$ to $0.6$~eV
and the single bands are not resolved.

%
%
%
%

Significant are the shifts in energy of both VB and CB.
Upon electron doping both the VB and the CB are shifted to higher energies 
whereas hole doping causes a shift to lower energies.
The gap between the VB and the CB has its maximum at zero charge and
decreases for positive as well as for negative charging by $0.1$--$0.2$~eV.
The gap between the VB and the states lower in energy as
well as the gap between the CB and the states higher in energy are also
changed.
In most of the cases the gap between the CB and the next higher bands is larger
than $0.6$~eV.
However, for both structures of \csixbr\ and $Q = -3$ this gap is only
$0.2$~eV. 
For each structure of \csixtx\ (X=Cl,Br) the gap between the VB and
the next lower states is almost constant upon electron doping whereas for
hole doping the VB are shifted towards lower energy and the next lower states
towards higher energy.
For \poneht\ \csixbr\ and $Q = +3$ the former single band below the 
VB has merged with the latter, i.e.~the gap has vanished.
The VB is formed of six instead of five states.
However, there is still a small gap of $\sim 0.1$~eV between these six bands
and the next lower states.
More drastic are the changes in the case of \pbonelt\ 
\csixbr\ and $Q = +3$ where the VB and the next lower states form a 
continuum.

\begin{table}
\caption{
Total density of states (DOS) at the Fermi energy of
charged pure fcc \csixty\ and \csixtx\ (X=Cl,Br).
All values are in states per unit cell and eV.
}
\label{tab:charges-and-dos}
\begin{ruledtabular}
\begin{tabular}{lrrrr}
 $Q$~\tbn{a} & +3 & +1 & -1 & -3 \\
\hline
fcc \csixty~\tbn{b}                                & 29 & 38 & 16 & 17 \\
\csixcl \\
\hspace{3mm} \poneht~\tbn{c}                       & 29 & 24 & 17 & 17 \\
\hspace{3mm} \pbonelt~\tbn{e}                      & 33 & 15 & 14 & 19 \\
 \cline{2-5}
\hspace{3mm} average~\tbn{f}                       & 31 & 20 & 16 & 18 \\
 & \\
\hspace{3mm} \poneht, without \clform~\tbn{c),d}   & 29 & 25 & 20 & 16 \\
 & \\
\csixbr \\
\hspace{3mm} \poneht~\tbn{c}                       & 34 & 31 & 29 & 17 \\
\hspace{3mm} \pbonelt~\tbn{e}                      & 37 & 19 & 24 & 29 \\
 \cline{2-5}
\hspace{3mm} average~\tbn{f}                       & 36 & 25 & 28 & 23 \\
\end{tabular}
\end{ruledtabular}

\tbn{a}~Total charge per unit cell.
\tbn{b}~Crystal structure of David \cite{david46-1995}.
\tbn{c}~Crystal structure of Jansen and Waidmann
        \cite{jansenwaidmann14-1995}.
\tbn{d}~The two \clform\ molecules are simply deleted from the unit cell
        of the \poneht\ structure and the remaining \csixty\ molecule
        is re-optimized.
\tbn{e}~Crystal structure of Dinnebier \etal\
        \cite{dinnebiergunnarssonbrummkochhuqstephensjansen-2002}.
\tbn{f}~Average of \poneht\ and \pbonelt.
\hfil
\end{table}

Table~\ref{tab:charges-and-dos} summarizes the calculated total DOS at the 
Fermi energy of charged pure and haloform intercalated \csixty\ for 
$Q = \pm 1$ and $Q = \pm 3$.
The accuracy of the calculated DOS is about $\pm 1$~state per unit cell and eV.

There is a clear trend that negatively charged systems have 
smaller DOS than positively charged systems.
The small differences in shapes of the DOS of the \poneht\ and 
\pbonelt\ structures (Fig.~\ref{fig:charges-and-tdos}) become apparent in 
remarkable differences between DOS at the Fermi energy.
The DOS at \Efermi\ of the \poneht\ structures are for some charges
larger and for other charges smaller than that of the \pbonelt\
structures.
The discrepancies are up to 12~states per eV and unit cell and are
primarily a result of different intermolecular distances
(Table \ref{tab:calc-structs}) caused by differently oriented haloform and 
\csixty\ molecules.
Furthermore, due to narrow VB and CB peaks the DOS is sensitive with respect to
the position of the Fermi energy.

%
However, at finite temperature the \poneht\ as well as the \pbonelt\
configuration can occur and the DOS at the Fermi energy for a particular 
charging would be an average of both
\cite{boltzmann}.
Except for $Q = +1$ the averaged DOS is increasing as
DOS(fcc \csixty) $<$ DOS(\csixcl) $<$ DOS(\csixbr)
(Table~\ref{tab:charges-and-dos}).
But also for $Q = +1$ \csixcl\ has on average a smaller DOS than \csixbr.

Table~\ref{tab:charges-and-dos} includes also the DOS at the Fermi energy
of the hypothetical hexagonal lattice of pure \csixty\ in the
geometry of \poneht\ \csixcl\ 
(see Section~\ref{subsubsec:bands}).
For all charges the calculations yields DOS that are very similar to that of
\poneht\ \csixcl.
Hence, the orbitals of the \clform\ molecules which interact
with the $\pi$ system of the \csixty\ subsystem have a
negligibly influence on the DOS of the entire charged system.

\section{Summary and conclusions}
\label{sec:concl}

Using density functional methods we have determined the electronic
properties (band structure, partial and total density of states,
charge distribution) of pure \csixty\ and \csixtx\ (X=Cl,Br) after
optimization of their atomic structures.
The changes resulting from doping with one and three electrons/holes
per unit cell have also been examined.

The calculations yielded for un-doped chloroform and bromoform intercalated 
\csixty, \csixtx\ (X=Cl,Br) indirect gaps between the
valence bands (VB) and conduction bands (CB) larger than 1~eV.
Both \csixcl\ and \csixbr\ are narrow band materials,
i.e.~most of the electronic states are mainly localized on the molecules.
The calculated width of the VB and CB are $0.4$--$0.6$~eV and $0.3$--$0.4$~eV,
respectively.
The orbitals of the haloform molecules have a considerable overlap
between the $\pi$ orbitals of the fullerene molecules and the $p$-type orbitals
of halogen atoms significantly contribute to the VB and CB of \csixtx.
Both compounds are mainly stabilized by dipole-induced dipole
interactions rather than by van-der-Waals interactions.
The intrinsically differently charged atoms of the haloform molecules cause a
polarization of the fullerene molecules.

Doping with charge carriers turns the \csixtx\ (X=Cl,Br) to metals.
As a result of the intermolecular overlap of the CHX$_3$ and \csixty\
orbitals, 10 to 20~\% of the charge of the doped systems
is on the haloform molecules instead of being completely localized on the
fullerene molecules.
Charging with electrons leads to a shift of both VB and CB
to higher energies whereas the doping with holes shift them to lower
energies.
The corresponding band widths undergo relatively small changes.
At a charging of $Q= +3$  per \csixtx\ the gap between the
VB and the states just below the VB is significantly reduced.
In the case of \csixbr\ the VB has even merged with states lower in energy.
The calculated density of states at the Fermi energy for different chargings
are clearly smaller for electron than for hole doping and are larger for
\csixbr\ than for \csixcl.
Calculations on different crystal structures of \csixcl\ and \csixbr\
revealed that the density of states at the Fermi energy are
sensitive to the orientation of the haloform and \csixty\ molecules.
At a given charge the differences between the various crystal structures
are up to 12~states per eV and unit cell.

At a charging of $Q = +3$, which resembles the superconducting phase of
pure \csixty\ and \csixtx\ (X=Cl,Br), we calculated DOS at the Fermi energy 
that increases as DOS(\csixty) $<$ DOS(\csixcl) $<$ DOS(\csixbr).
On the other hand, the tight-binding calculations of Dinnebier \etal\
\cite{dinnebiergunnarssonbrummkochhuqstephensjansen-2002} 
yielded DOS that have the reverse order.
This discrepancy is related to the different methods considered.
The method used in the present work is more accurate than the
tight-binding approach.
Nevertheless, our calculations clearly supports the conclusion of 
Dinnebier \etal\ 
\cite{dinnebiergunnarssonbrummkochhuqstephensjansen-2002} that the
DOS alone cannot account for the observed increase of $T_c$ upon \clform\
intercalation and substitution of Cl by Br.
According to a different scenario proposed by Bill and Kresin
\cite{billkresin26-2002} the increase of $T_c$ upon intercalation is caused
by an additional contribution to the superconducting pairing interaction
which comes from the coupling of charge 
carriers with the vibrational manifold of the intercalated molecules.
The difference in vibrational spectra of \clform\ and 
\brform\ molecules \cite{herzberg-1964} leads to a noticeable increase of 
$T_c$ upon Cl$\rightarrow$Br substitution.

The present calculations have all been performed on a three-dimensional crystal
in absence of an applied electric field.
However, the presence of an electric field may lead to a re-orientation and
polarization of the haloform and fullerene molecules.
Such structural changes affect the electronic structure of the system and may
influence vibronic couplings as well.
It would be of interest to complement the study presented
in this article with calculations on \csixtx\ (X=Cl,Br) in an
electric field.
A similar problem has been examined for pure \csixty\
\cite{wehrlipoilblancrice23-2001}.

%

%
\begin{acknowledgments}
We thank P.W.~Stephen for sending the manuscript
of Ref.~\cite{dinnebiergunnarssonbrummkochhuqstephensjansen-2002}
prior to publication.
\end{acknowledgments}

%


\begin{thebibliography}{25}
\expandafter\ifx\csname natexlab\endcsname\relax\def\natexlab#1{#1}\fi
\expandafter\ifx\csname bibnamefont\endcsname\relax
  \def\bibnamefont#1{#1}\fi
\expandafter\ifx\csname bibfnamefont\endcsname\relax
  \def\bibfnamefont#1{#1}\fi
\expandafter\ifx\csname citenamefont\endcsname\relax
  \def\citenamefont#1{#1}\fi
\expandafter\ifx\csname url\endcsname\relax
  \def\url#1{\texttt{#1}}\fi
\expandafter\ifx\csname urlprefix\endcsname\relax\def\urlprefix{URL }\fi
\providecommand{\bibinfo}[2]{#2}
\providecommand{\eprint}[2][]{\url{#2}}

\bibitem[{\citenamefont{{M. Jansen and G. Waidmann}}(1995)}]{jansenwaidmann14-1995}
\bibinfo{author}{\bibnamefont{{M. Jansen and G. Waidmann}}}, \bibinfo{journal}{Z. Anorg.
  Allg. Chem.} \textbf{\bibinfo{volume}{14}}, \bibinfo{pages}{621}
  (\bibinfo{year}{1995}).

\bibitem[{\citenamefont{{C. Collins, J. Foulkes, A. D. Bond and J. Klinowski
}}(1999)}]{collinsfoulkesbondklinowski1-1999}
\bibinfo{author}{\bibnamefont{{C. Collins, J. Foulkes, A. D. Bond and
  J. Klinowski}}}, \bibinfo{journal}{Phys. Chem. Chem. Phys.}
  \textbf{\bibinfo{volume}{1}}, \bibinfo{pages}{5323} (\bibinfo{year}{1999}).

\bibitem[{\citenamefont{{R. E. Dinnebier, O. Gunnarsson, H. Brumm, E. Koch,
  A. Huq, P. W. Stephens and M. Jansen
}}(2002)}]{dinnebiergunnarssonbrummkochhuqstephensjansen-2002}
\bibinfo{author}{\bibnamefont{{R. E. Dinnebier, O. Gunnarsson, H. Brumm,
  E. Koch, A. Huq, P. W. Stephens and M. Jansen}}}, \bibinfo{journal}{Science}
  \textbf{\bibinfo{volume}{296}}, \bibinfo{pages}{109} (\bibinfo{year}{2002}).

\bibitem[{\citenamefont{{C. Collins, M. Duer and J. Klinowski
}}(2000)}]{collinsduerklinowski321-2000}
\bibinfo{author}{\bibnamefont{{C. Collins, M. Duer and J. Klinowski}}},
  \bibinfo{journal}{Chem. Phys. Lett.} \textbf{\bibinfo{volume}{321}},
  \bibinfo{pages}{287} (\bibinfo{year}{2000}).

\bibitem[{\citenamefont{{A. Hebard}}(1992)}]{hebard45-1992}
\bibinfo{author}{\bibnamefont{{A. Hebard}}}, \bibinfo{journal}{Phys. Today}
  \textbf{\bibinfo{volume}{45}}, \bibinfo{pages}{26} (\bibinfo{year}{1992}).

\bibitem[{\citenamefont{{O. Gunnarson}}(1997)}]{gunnarson69-1997}
\bibinfo{author}{\bibnamefont{{O. Gunnarson}}}, \bibinfo{journal}{Rev. Mod.
  Phys.} \textbf{\bibinfo{volume}{69}}, \bibinfo{pages}{575}
  (\bibinfo{year}{1997}).

\bibitem[{\citenamefont{{O. Gunnarson, E. Koch and R. Martin
}}(1998)}]{gunnarsonkochmartin-1998}
\bibinfo{author}{\bibnamefont{{O. Gunnarson, E. Koch and R. Martin}}}, in
  \emph{\bibinfo{booktitle}{{Pair} {Correlations} in {Many-Fermion}
  {Systems}}}, edited by \bibinfo{editor}{\bibfnamefont{V.~Z.}
  \bibnamefont{Kresin}} (\bibinfo{publisher}{Plenum}, \bibinfo{address}{New
  York}, \bibinfo{year}{1998}), p. \bibinfo{pages}{155}.

\bibitem[{\citenamefont{{J. H. Sch\"on, C. Kloc and B. Batlogg
}}(2000)}]{schoenklocbatlogg408-2000}
\bibinfo{author}{\bibnamefont{{J. H. Sch\"on, C. Kloc and B. Batlogg}}},
  \bibinfo{journal}{Nature} \textbf{\bibinfo{volume}{408}},
  \bibinfo{pages}{549} (\bibinfo{year}{2000}).

\bibitem[{\citenamefont{{A. Ramirez}}()}]{ramirez-june2002}
\bibinfo{author}{\bibnamefont{{A. Ramirez}}}, \bibinfo{howpublished}{personal
  communication}.

\bibitem[{\citenamefont{{R. F. Service}}(2002)}]{service296-2002}
\bibinfo{author}{\bibnamefont{{R. F. Service}}}, \bibinfo{journal}{Science}
  \textbf{\bibinfo{volume}{296}}, \bibinfo{pages}{1584} (\bibinfo{year}{2002}).

\bibitem[{\citenamefont{{J. H. Sch\"on, C. Kloc and B. Batlogg
}}(2001)}]{schoenklocbatlogg293-2001}
\bibinfo{author}{\bibnamefont{{J. H. Sch\"on, C. Kloc and B. Batlogg}}},
  \bibinfo{journal}{Science} \textbf{\bibinfo{volume}{293}},
  \bibinfo{pages}{2432} (\bibinfo{year}{2001}).

\bibitem[{\citenamefont{{F. Meunier, J. P. Burger, G. Deutscher and E. Guyon
}}(1968)}]{meunierburgerdeutscherguyon26-1968}
\bibinfo{author}{\bibnamefont{{F. Meunier, J. P. Burger, G. Deutscher and E. Guyon
}}}, \bibinfo{journal}{Phys. Lett.} \textbf{\bibinfo{volume}{26 A}},
  \bibinfo{pages}{309} (\bibinfo{year}{1968}).

\bibitem[{\citenamefont{{V. Z. Kresin}}(1974)}]{kresin49-1974}
\bibinfo{author}{\bibnamefont{{V. Z. Kresin}}}, \bibinfo{journal}{Phys.
  Lett.} \textbf{\bibinfo{volume}{49 A}}, \bibinfo{pages}{117}
  (\bibinfo{year}{1974}).

\bibitem[{\citenamefont{{W. I. F. David}}(1995)}]{david46-1995}
\bibinfo{author}{\bibnamefont{{W. I. F. David}}}, \bibinfo{journal}{Appl.
  Radiat. Isotopes} \textbf{\bibinfo{volume}{46}}, \bibinfo{pages}{519}
  (\bibinfo{year}{1995}).

\bibitem[{\citenamefont{{B. Delley}}(1990)}]{delley921-1990}
\bibinfo{author}{\bibnamefont{{B. Delley}}}, \bibinfo{journal}{J. Chem. Phys.}
  \textbf{\bibinfo{volume}{92}}, \bibinfo{pages}{508} (\bibinfo{year}{1990}).

\bibitem[{\citenamefont{{B. Delley}}(2000)}]{delley11318-2000}
\bibinfo{author}{\bibnamefont{{B. Delley}}}, \bibinfo{journal}{J. Chem. Phys.}
  \textbf{\bibinfo{volume}{113}}, \bibinfo{pages}{7756} (\bibinfo{year}{2000}).

\bibitem[{\citenamefont{{A. D. Becke}}(1993)}]{becke987-1993}
\bibinfo{author}{\bibnamefont{{A. D. Becke}}}, \bibinfo{journal}{J. Chem.
  Phys.} \textbf{\bibinfo{volume}{98}}, \bibinfo{pages}{5648}
  (\bibinfo{year}{1993}).

\bibitem[{\citenamefont{{J. P. Perdew and Y. Wang}}(1992)}]{perdewwang4523-1992}
\bibinfo{author}{\bibnamefont{{J. P. Perdew and Y. Wang}}},
  \bibinfo{journal}{Phys. Rev. B} \textbf{\bibinfo{volume}{{45}}},
  \bibinfo{pages}{13244} (\bibinfo{year}{1992}).

\bibitem[{\citenamefont{{J. Andzelm, R. D. King-Smith and G. Fitzgerald
}}(2001)}]{andzelmkingsmithfitzgerald335-2001}
\bibinfo{author}{\bibnamefont{{J. Andzelm, R. D. King-Smith and G. Fitzgerald
}}}, \bibinfo{journal}{Chem. Phys. Lett.} \textbf{\bibinfo{volume}{335}},
  \bibinfo{pages}{321} (\bibinfo{year}{2001}).

\bibitem[{\citenamefont{{D. R. Lide}}(1998-1999)}]{handbookchemphys-1999}
\bibinfo{author}{\bibnamefont{{D. R. Lide}}}, \emph{\bibinfo{title}{{Handbook}
  of {Physics} and {Chemistry}}} (\bibinfo{publisher}{CRC Press},
  \bibinfo{address}{{Boca Raton}}, \bibinfo{year}{1998-1999}),
  \bibinfo{edition}{79th} ed.

\bibitem[{\citenamefont{{A. Ruiz, J. Bret\'on and J. M. Gomez Llorente
}}(2001)}]{ruizbretongomezllorente1143-2001}
\bibinfo{author}{\bibnamefont{{A. Ruiz, J. Bret\'on and J. M. Gomez Llorente
}}}, \bibinfo{journal}{J. Chem. Phys.} \textbf{\bibinfo{volume}{114}},
  \bibinfo{pages}{1272} (\bibinfo{year}{2001}).

\bibitem[{\citenamefont{{M. J. Winter.}}(2002)}]{webelements}
\bibinfo{author}{\bibnamefont{{M. J. Winter}}},
  \bibinfo{howpublished}{WebElements$^{\rm TM}$, the periodic table on the WWW,
  http://www.webelements.com/} (\bibinfo{year}{2002}).

\bibitem[{\citenamefont{{R. W. Lof, M. A. van Veenendaal, B. Koopmans,
H. T. Jonkman and G. A. Sawatzky}}(1992)}]
 {lofveenendaalkoopmansjonkmansawatzky6826-1992}
\bibinfo{author}{\bibnamefont{{R. W. Lof, M. A. van Veenendaal,
B. Koopmans, H. T. Jonkman and G. A. Sawatzky}}},
\bibinfo{journal}{Phys. Rev. Lett.} \textbf{\bibinfo{volume}{68}},
\bibinfo{pages}{3924} (\bibinfo{year}{1992}).

\bibitem[{\citenamefont{{J. C. Slater}}(1972)}]{slater6-1972}
\bibinfo{author}{\bibnamefont{{J. C. Slater}}},
\bibinfo{journal}{Adv. Quantum Chem.} \textbf{\bibinfo{volume}{6}},
  \bibinfo{pages}{1} (\bibinfo{year}{1972}).

\bibitem[{\citenamefont{{D. R. Liberman}}(1999)}]{liberman6211-1999}
\bibinfo{author}{\bibnamefont{{D. A. Liberman}}},
\bibinfo{journal}{Phys. Rev. B} \textbf{\bibinfo{volume}{62}},
 \bibinfo{pages}{6851} (\bibinfo{year}{1999}).

\bibitem[{\citenamefont{{F. L. Hirshfeld}}(1977)}]{hirshfeld44-1977}
\bibinfo{author}{\bibnamefont{{F. L. Hirshfeld}}},
  \bibinfo{journal}{Theor. Chim. Acta} \textbf{\bibinfo{volume}{44}},
  \bibinfo{pages}{129} (\bibinfo{year}{1977}).
  \bibinfo{note}{The charge of an atom $N$ is calculated as
$q(N) = \int \rho_d(\bdr) W_N(\bdr) d\bdr$, where
$W_N(\bdr) = \rho_N(\bdr - \bdR_N) [\sum_M \rho_M(\bdr - \bdR_M)]^{-1}$
is the weight and $M$ is the sum index over {\it all} atoms.
The deformation charge density, $\rho_d$, is the difference between the charge
density of the system and the density of the separated atoms and is
given by
$\rho_d(\bdr) = \rho(\bdr) - \sum_M \rho_M(\bdr - \bdR_M)$.}

\bibitem[{\citenamefont{{Boltzmann}}(2002)}]{boltzmann}
  \bibinfo{note}{
At a temperature of $T = 100$~K the Boltzmann statistic yields for the ratio 
of the occurrence of the \poneht\ and the \pbonelt\ configurations of \csixcl\
almost one when the energy difference between both configuration is
28~\kjmol\ (see Table~\ref{tab:calc-structs}).
}

\bibitem[{\citenamefont{{A. Bill and V. Z. Kresin}}(2002)}]{billkresin26-2002}
\bibinfo{author}{\bibnamefont{{A. Bill and V. Z. Kresin}}},
  \bibinfo{journal}{Eur. Phys. J. B} \textbf{\bibinfo{volume}{26}},
  \bibinfo{pages}{3} (\bibinfo{year}{2002}).

\bibitem[{\citenamefont{{G. Herzberg}}(1964)}]{herzberg-1964}
\bibinfo{author}{\bibnamefont{{G. Herzberg}}},
  \emph{\bibinfo{title}{{Molecular} {Spectra} and {Molecular} {Structure}. II.
  {Infrared} and {Raman} {Spectra} of {Polyatomic} {Molecules}}}
  (\bibinfo{publisher}{D. Van Nostrad}, \bibinfo{address}{Princeton},
  \bibinfo{year}{1964}).

\bibitem[{\citenamefont{{S. Wehrli, D. Poilblanc and T. Rice
}}(2001)}]{wehrlipoilblancrice23-2001}
\bibinfo{author}{\bibnamefont{{S. Wehrli, D. Poilblanc and T. Rice}}},
  \bibinfo{journal}{Eur. Phys. J. B} \textbf{\bibinfo{volume}{23}},
  \bibinfo{pages}{345} (\bibinfo{year}{2001}).

\end{thebibliography}
\end{document}